# Neutrinos of non-zero rest mass and the equivalence principle

**D. G. Banhatti**
Radio Astronomy Centre (TIFR), P O Box 8, Ootacamund 643001, India
and
TIFR Centre for Radio Astronomy, P O Box 1234, Bangalore 560012, India

**Abstract.** Assuming that neutrinos of non-zero rest mass dominate the mass density in the universe, and also the mass density on the scale of clusters of galaxies, one obtains the upper limit $m <\approx 20$ eV/c$^2$ on their mass, independent of the values of $H_0$ and $q_0$, and the lower limit $m >\approx 5$ eV/c$^2$ independent of $q_0$ and almost independent of $H_0$. If they are <u>not</u> to dominate the mass density on the scale of binary galaxies and small groups of galaxies, one must have $m <\approx 14$ eV/c$^2$ independent of $q_0$ and almost independent of $H_0$. Going one step further, we allow neutrinos to have different gravitational and inertial masses so that r = gravitational / inertial mass. Then using m for the inertial mass, we have $m.r^{1/4} >\approx 5$ eV/c$^2$, $m.r^{1/4} <\approx 14$ eV/c$^2$ and $m.r <\approx 20$ eV/c$^2$, which together imply $r <\approx 6.3$. For a specific value, say, 12 eV/c$^2$, for m, we have $0.03 <\approx r <\approx 1.7$.

**Introduction**
The possibility that neutrinos may have non-zero rest mass has led to the investigation of their possible role in the dynamics of astrophysical systems in the universe, including their effect on the dynamics of the universe itself (Schramm & Steigman 1981a, b; Tremaine & Gunn 1979). If neutrinos dominate the mass density in the universe, estimates of the age of the universe from nucleocosmochronometry put an upper limit on the mass m of the neutrino (Joshi & Chitre 1981a, b). If they dominate the mass density on the scale of clusters of galaxies also, one gets a lower limit on the mass (Tremaine & Gunn 1979). The assumption that they do <u>not</u> dominate the mass in binaries and small groups of galaxies leads to an upper limit slightly less than that obtained from the age of the universe. Allowing the neutrino to have different gravitational and inertial masses, we assume that it does not necessarily obey the equivalence principle and let r denote the ratio of its gravitational to inertial mass. Incorporating the possibility that $r \neq 1$ in the calculation for the (inertial) mass m spreads the three limits into a region in the r-m plane.

**The three limits**
<u>Upper limit from age of universe</u>
Writing the total mass density $\rho_0$ as the sum of the non-ν mass density $\rho_m$ and the ν mass density $\rho_{G\nu}$,

$$\rho_0 = \rho_m + \rho_{G\nu} = \rho_m + r.\rho_\nu.$$

Subscript G refers to gravitational mass and $\rho_{Gv} = r.\rho_v$ ($\rho_v$ being the inertial mass density) to allow for different coupling between gravitational and inertial masses for neutrinos as compared with non-$v$ matter. If neutrinos dominate,

$$\rho_0 \approx r.\rho_v = r.m.n_v$$

where $n_v$ is the number density of the three presently known types of neutrinos and m is the inertial mass of the neutrino (assumed same for all three types). Inserting this in the upper limit to the age of the universe as derived for both the Friedmann world models (Joshi & Chitre 1981a) and for the generally hyperbolic world models (Pankaj Joshi, private communication), of which the Friedmann models are a special case,

$$t_{0max} = \sqrt{(A / G. \rho_0)} \approx \sqrt{(A / G. r.m.n_v)};$$

$A = 3\pi / 32$ for Friedmann models, and $3\pi / 16$ for general globally hyperbolic world-models. Note that this limit is independent of the values of $H_0$ and $q_0$. (Since the limit for the Friedmann models is tighter, we use that for numerical calculations below.) Let $t_U$ be the maximum of the various estimates of the age of the universe obtained from the analysis of relative isotope abundances, helium abundance, dynamical considerations for globular clusters, etc. $t_U$ is thus a lower limit on the age of the universe, independent of $H_0$- and $q_0$-values (Symbalisty et al 1980). Therefore, $t_{0max} > t_U$, which implies

$$\begin{aligned} r.m &< A / G.t_U^2.n_v, \\ &= 3\pi / 32. G.t_U^2.n_v \quad \text{for Friedmann models.} \end{aligned} \quad (1)$$

Lower limit from clusters of galaxies
Examining the missing light problem on the scales of various astrophysical systems, Schramm & Steigman (1981a, b) found that its severity increases with size scale of the system (from galaxies to binaries to small groups to clusters). This, together with the ease with which heavier neutrinos can collapse on smaller scales compared to lighter neutrinos led them to conclude that non-$v$ matter (nucleons) can account for the mass density on all scales smaller than clusters of galaxies. Relic neutrinos could have collapsed on the scale of clusters of galaxies after they had cooled sufficiently, provided the gravitational potentials of the clusters were deep enough. A necessary condition for this collapse is that the maximum value of the phase-space density decreases in the transition from a free Fermi distribution (at a temperature of $\approx 1$ MeV/k) to an isothermal distribution (at present) (Tremaine & Gunn 1979). We write $r.\rho_v$ instead of $\rho_v$ in the calculation of Tremaine & Gunn to obtain

$$r.m^4 > 9.h^3 / (2\pi)^{5/2}.N.g_v.G.\sigma_{cl}.R_{cl}^2 \quad (2)$$

where N is the number of species of neutrinos ($v$ and $v$-bar counted as two different species), $g_v$ is the number of helicity states (assumed the same for all species) and $\sigma_{cl}$ and $R_{cl}$ are the one-dimensional velocity dispersion and core radius of the typical cluster of galaxies.
---------------------------------------------------------------------------

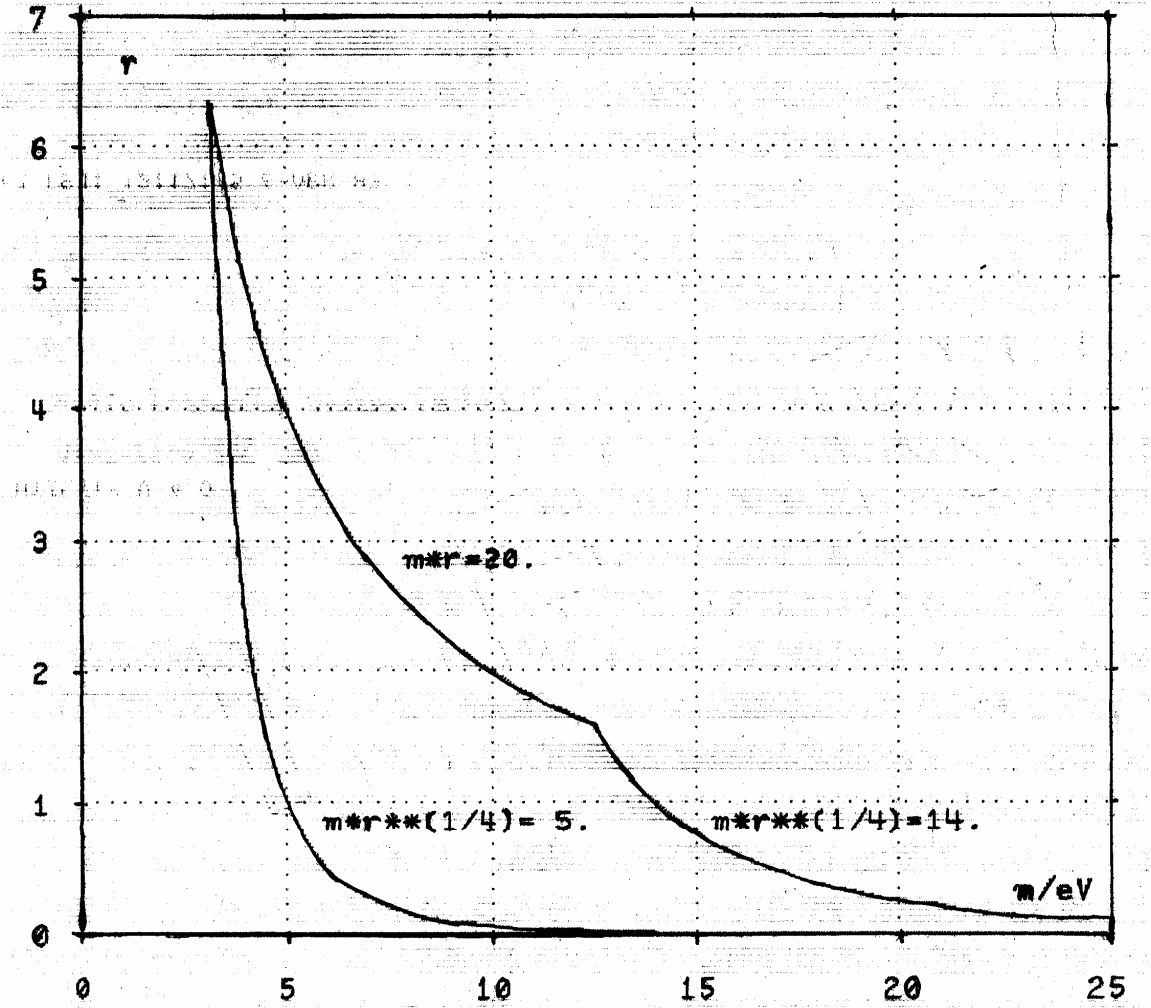

**Fig.1.** r ≡ gravitational / inertial mass of the neutrino, and m ≡ inertial mass of the neutrino. The allowed region in the r-m plane is shown.

---------------------------------------------------------------

Upper limit from binaries and small groups
Coming down in size from clusters of galaxies, the next smaller astrophysical systems are binary galaxies and small groups of galaxies. Applying the same principles as above, if neutrinos are <u>not</u> to collapse on the scale of binary galaxies and small groups of galaxies, inequality (2) is reversed, with $\sigma_{B,SG}$ and $R_{B,SG}$ the typical relative velocity and separation between members in these astrophysical systems:

$$r.m^4 < 9.h^3 / (2\pi)^{5/2}.N.g_v.G.\sigma_{B,SG}.R_{B,SG}^2 \tag{3}$$

**Numerical results**
We now put numerical values in (1)-(3). With $t_U = 20$ Gyr and $n_v = 300$ per cc in (1),

$$r.m < 20 \text{ eV}/c^2; \tag{1'}$$

and taking N = 6 species of neutrinos (the e, μ and τ neutrinos and antineutrinos), each with $g_\nu$ = 2 helicity states, (2) gives, for $\sigma_{cl}$ = $10^3$ km/sec, $R_{cl}$ = 250.(50/$H_0$) kpc, for the typical cluster of galaxies,

$r^{1/4}.m > 5.\sqrt{(H_0/50)}$ eV/$c^2$. (2')

Substituting $\sigma_{B,SG}$ = 100 km/sec and $R_{B,SG}$ = 100 kpc in (3) as the typical orbital velocity and separation between members of binary galaxies and small groups of galaxies, we get

$r^{1/4}.m < 14.\sqrt{(H_0/50)}$ eV/$c^2$. (3')

The region delimited by (1'), (2') and (3') in the r-m plane is shown in Fig.1. From the diagram, r <≈ 6.3 independent of the value of m. For a specific value, say, 12 eV/$c^2$, for m, we have 0.03 <≈ r <≈ 1.7.

Thus, we see that the dynamics of the universe and of clusters of galaxies, if dominated by the neutrino, together with an estimate of the age of the universe from non-cosmological considerations and the assumption that neutrinos of non-zero rest mass are <u>not</u> necessary for collapse on a small enough scale (viz, binaries and small groups of galaxies), are consistent with the neutrino satisfying the equivalence principle. This conclusion is independent of the value of the cosmological deceleration parameter $q_0$ and only weakly dependent on that of the Hubble parameter $H_0$. Should direct estimates of the core radius of a nearby cluster of galaxies and of the separation between members of small groups of galaxies become available, (2') and (3') will no longer have a dependence on $H_0$.

**Acknowledgments**
This work was spurred by the idea of making the calculations of Prof Arthur Halprin (on similar lines (private communication) $H_0$- and $q_0$-independent. I also thank Pankaj Joshi for encouragement and communicating his latest unpublished results on the general globally-hyperbolic world-models. It is a pleasure to thank Chris Salter for moral support and for reading the manuscript.